\documentclass[conference]{IEEEtran}
\IEEEoverridecommandlockouts
\usepackage{algorithm}
\usepackage[noend]{algpseudocode}
\usepackage{booktabs}
\usepackage{cite}
\usepackage{amsmath,amssymb,amsfonts}
\usepackage{csquotes}
\usepackage{graphicx}
\usepackage{hyperref}
\usepackage[autolanguage]{numprint}
\usepackage{textcomp}
\usepackage{tikz}
\usetikzlibrary{patterns}
\usetikzlibrary{positioning}
\usepackage{subcaption}
\usepackage{xcolor}
\usepackage{xspace}
\usepackage{balance}

\newcommand{\ie}{i.\,e.,\xspace}
\newcommand{\eg}{e.\,g.,\xspace}
\newcommand{\wrt}{w.\,r.\,t.\xspace}

\newcommand{\etal}{et al.\xspace}

\newcommand{\bc}{\mathbf{b}}
\newcommand{\bccnt}{\mathbf{\widetilde{c}}}
\newcommand{\bcapprox}{\mathbf{\widetilde{b}}}
\newcommand{\loc}{\mathrm{loc}}

\newcommand{\Oh}{\ensuremath{\mathcal{O}}}

\newcommand{\realTotalSu}{\numprint{7.4}}
\newcommand{\realAdsSu}{\numprint{16.1}}

\begin{document}

\title{Scaling Betweenness Approximation to Billions of Edges by MPI-based Adaptive Sampling
\thanks{This work is partially supported by German Research Foundation (DFG) grant ME 3619/3-2
within Priority Programme 1736 \textit{Algorithms for Big Data}.
}
}

\author{\IEEEauthorblockN{Alexander van der Grinten}
\IEEEauthorblockA{\textit{Department of Computer Science} \\
\textit{Humboldt-Universit\"at zu Berlin}\\
Berlin, Germany \\
avdgrinten@hu-berlin.de}
\and
\IEEEauthorblockN{Henning Meyerhenke}
\IEEEauthorblockA{\textit{Department of Computer Science} \\
\textit{Humboldt-Universit\"at zu Berlin}\\
Berlin, Germany \\
meyerhenke@hu-berlin.de}
}

\maketitle

\pagestyle{plain}

\begin{abstract}
Betweenness centrality is one of the most popular vertex centrality measures in network analysis. Hence, many (sequential and parallel) algorithms to compute or approximate betweenness have been devised. Recent algorithmic advances have made it possible to approximate betweenness very efficiently on shared-memory architectures. Yet, the best shared-memory algorithms can still take hours of running time for large graphs, especially for graphs with a high diameter or when a small relative error is required.

In this work, we present an MPI-based generalization of the state-of-the-art shared-memory algorithm for betweenness approximation. This algorithm is based on adaptive sampling; our parallelization strategy can be applied in the same manner to adaptive sampling algorithms for other problems. In experiments on a 16-node cluster, our MPI-based implementation is by a factor of 16.1x faster than the state-of-the-art shared-memory implementation when considering our parallelization focus -- the adaptive sampling phase -- only. For the complete algorithm, we obtain an average (geom. mean) speedup factor of 7.4x over the state of the art. For some previously very challenging inputs, this speedup is much higher. As a result, our algorithm is the first to approximate betweenness centrality on graphs with several billion edges in less than ten minutes with high accuracy.

\end{abstract}


\section{Introduction}
Betweenness centrality (BC) is a popular vertex centrality measure in network analysis.
As such, it assigns a numerical score to each vertex of a graph to
quantify the importance of the vertex.
In particular, the (normalized) betweenness $\bc(x)$ of a vertex $x \in V$ in a graph $G = (V, E)$
is defined as $\bc(x) := \frac1{n(n-1)} \sum_{s \neq t} \frac{\sigma_{st}(x)}{\sigma_{st}}$,
where $\sigma_{st}(x)$ is the number of shortest $s$-$t$ paths over $x$
and $\sigma_{st}$ is the total number of shortest $s$-$t$ paths.
Betweenness has applications in many domains: to name just a few, Bader \etal~\cite{bader2007approximating}
mention lethality in biological networks, study of sexually transmitted
diseases, identifying key actors in terrorist networks, organizational behavior, and supply chain management processes.
BC is, however, known to be expensive to compute.
The classical exact algorithm by Brandes~\cite{doi:10.1080/0022250X.2001.9990249} has a running time of $O(|V| |E|)$;
it is still the basis for many exact parallel or distributed algorithms today.
Moreover, recent theoretical results suggest that
it is unlikely that a subcubic algorithm (\wrt $|V|$) exists
to compute betweenness exactly on arbitrary graphs~\cite{DBLP:conf/soda/AbboudGW15}.
Thus, for large graphs beyond, say, $5$M vertices and $100$M edges, exact betweenness algorithms
can be seen as hardly practical, not even parallel and distributed ones (see Section~\ref{sec:relwork}).

% Current SotA
For betweenness \emph{approximation}, however, fast practical algorithms
exist. The fastest known betweenness approximation algorithm is the
KADABRA algorithm by Borassi and Natale~\cite{DBLP:conf/esa/BorassiN16}.
This algorithm uses sampling to obtain a probabilistic
guarantee on the quality of the approximation:
with a probability of $(1 - \delta)$, the resulting betweenness
values never deviate by more than an additive error of
$\pm \epsilon$ from the true values,
for all vertices of the input graph.
$\delta$ and $\epsilon$ are constants that can be chosen arbitrarily
(but smaller choices increase the algorithm's running time).
More specifically, KADABRA performs adaptive sampling, \ie the algorithm
does \emph{not} compute a fixed number of samples \textit{a priori}.
Instead, the algorithm's stopping condition depends
on the samples that have been taken so far -- this
fact implies that parallelization is considerably more challenging
for adaptive sampling algorithms
than for \enquote{traditional} (\ie non-adaptive) sampling.
In particular, in the context of parallelization,
checking the stopping condition of adaptive sampling algorithms
requires some form of global aggregation of samples.

Despite these difficulties, a shared-memory parallelization of the KADABRA algorithm
was presented by the authors of this paper (among others)
in Ref.~\cite{DBLP:conf/europar/GrintenAM19}.
While this algorithm scales to graphs with hundreds
of millions of edges with an error bound of $\epsilon = 0.01$,
betweenness approximation for really
large graphs is still out of reach for current algorithms.
Especially when a higher accurary is desired,
current shared-memory algorithms quickly become impractical.
In fact, on many graphs only a handful of vertices
have a betweenness score larger than $0.01$ (\eg 38 vertices out of the
41 million vertices of the widely-studied \texttt{twitter} graph;
the situation is similar for other social networks and web graphs)
-- for $\epsilon = 0.01$, an approximation algorithm can only
reliably detect a small fraction of vertices
with highest betweenness score.
Choosing $\epsilon = 0.001$ improves this fraction of
reliably identified vertices
by an order of magnitude but requires multiple hours of
running time for large graphs on shared-memory machines.

\subsection{Scope and Challenges}
The goal of this paper is to provide an efficient MPI-based algorithm
for parallel betweenness approximation.
Inside each MPI process, we want to employ multithreading to utilize
modern multicore CPUs.
We consider the scenario in which a single sample can be taken locally,
\ie independently by each thread and in parallel, without involving any communication.
Checking the stopping condition,
however, does involve global synchronization
(\ie it requires a global aggregation of per-thread sampling states).
In this scenario, the main challenge is that we want to reduce the
time lost during communication to a minimum even though
we need to perform frequent global synchronization.

Note that the assumption that we can take samples locally
implies that we do not work with a distributed graph
data structure. This constrains our algorithm to input graphs that fit
into the memory of a single compute node. Fortunately, this is not an issue
for betweenness approximation: today's compute nodes usually have more than
enough memory to store the graphs for which betweenness computations
are feasible. In fact, we can fit \emph{all} networks from well-known
graph repositories like SNAP~\cite{DBLP:journals/tist/LeskovecS16}
and KONECT~\cite{DBLP:conf/www/Kunegis13} into the 96 GiB of RAM
that are available to MPI processes
on the compute nodes considered in our experiments.

\subsection{Our Capabilities}

In this paper, we present a new MPI-based algorithm for
betweenness approximation. Our algorithm is a parallelization of the
adaptive sampling algorithm KADABRA. As main techniques,
we use an efficient concurrent data structure from Ref.~\cite{DBLP:conf/europar/GrintenAM19}
to aggregate
sampling states from multiple threads, now combined with MPI reductions
to aggregate these sampling states across process boundaries.
We evaluate this new MPI algorithm on a cluster of 16 compute nodes.
The capabilities of our algorithm can be summarized as follows:
\begin{itemize}
	\item On a collection of 10 real-world data sets with up to $3.3$ billion edges,
		our MPI algorithm achieves a (geom.) mean speedup of \realTotalSu $\times$
		over the state-of-the-art shared-memory parallelization
		running on a single compute node.
	\item The focus of our parallelization is on the algorithm's
		adaptive sampling phase
		-- if we consider only this phase,
		we achieve a speedup of \realAdsSu $\times$.
		In particular, by avoiding NUMA-related bottlenecks, our MPI algorithm
		outperforms the state-of-the-art shared-memory implementation
		even on a single compute node by 20-30\%.
	\item Our code is by far the fastest betweenness approximation
	   available; it can handle graphs with a few billion edges in
	   less than ten minutes with an accuracy of $\epsilon = 0.001$.
\end{itemize}

%========================================================================================

\section{Related Work}
\label{sec:relwork}
A comprehensive overview on betweenness centrality algorithms is beyond the scope of this paper.
We thus focus on recent parallel approaches and in particular on parallel approximation algorithms for BC.

%%% "standard" algorithm
\paragraph*{Exact algorithms}
The algorithm with the fastest asymptotic time complexity for computing all BC values
is due to Brandes~\cite{doi:10.1080/0022250X.2001.9990249}. It performs $|V|$ augmented 
single-source shortest path (SSSP)
searches and requires $O(|V| |E|)$ time on unweighted graphs.
The main algorithmic idea is to express BC contributions by a sum over
a recursive formula. 
This recursion is evaluated after each SSSP search by accumulating the contributions
bottom-up in the corresponding SSSP tree.

Even if some graph manipulations (see \eg Ref.~\cite{DBLP:journals/tkdd/SariyuceKSC17})
help to speed up the Brandes approach,
major theoretical improvements \wrt the sequential time complexity (beyond fixed-parameter results) seem unlikely~\cite{DBLP:conf/soda/AbboudGW15}.
Of course, one can resort to parallelism for acceleration.
Consequently, there exist numerous parallel or distributed algorithms for computing BC exactly in various computational models:
shared-memory parallel~\cite{DBLP:conf/ipps/MadduriEJBC09}, 
parallel with distributed-memory~\cite{Solomonik:2017:SBC:3126908.3126971}, 
fully-distributed~\cite{hoang2019round} as well as (multi-)GPU systems~\cite{Sariyuce:2013:BCG:2458523.2458531,McLaughlin:2018:AGB:3241891.3230485,Bernaschi:2016:SBC:2903150.2903153} (among others).
Some of these papers operate on massive graphs -- in these cases they report running times for a relatively small sample
of SSSP searches only. Others report running times for the complete computation --
but then usually on graphs of moderate size only. We observe two exceptions:
Sar\i y\"uce \etal~\cite{Sariyuce:2013:BCG:2458523.2458531} provide exact results for graphs with
up to $234$M edges, but their computation on their largest instance alone requires more than 11 \emph{days}
on a heterogeneous CPU-GPU system. The second exception is due to AlGhamdi \etal~\cite{AlGhamdi:2017:BBC:3085504.3085510};
they also invest considerable supercomputing time ($2$M core hours) for providing BC scores on $20$ 
graphs (some of them very small) with up to $126$M edges.

\paragraph*{Approximation algorithms}
To deal with the ``$\Theta(|V||E|)$ complexity barrier'' in practice, several
approximation algorithms have been proposed. An empirical comparison between many of them
can be found in Matta \etal~\cite{Matta2019}.
Different from previous sampling-based approaches~\cite{brandes2007centrality,bader2007approximating,geisberger2008better}
(which we do not describe in detail due to space constraints, see~\cite{Matta2019} instead)
is the method by Riondato and Kornaropoulos~\cite{riondato2016fast}: 
it samples node \emph{pairs} (instead of SSSP sources)
and shortest paths between them. The algorithm, let us call it RK,
approximates the betweenness score of $v \in V$ as the fraction
of sampled paths that contain $v$ as intermediate node.
This approach yields a probabilistic absolute approximation guarantee on the solution quality:
the approximated BC values differ by at most $\epsilon$ from the exact values with probability at 
least $(1-\delta)$, where $\epsilon, \delta > 0$ can be arbitrarily small constants. 
%While such an absolute approximation works well for highly ranked nodes (as they have high scores),
%the relative position of nodes with low ranks should be treated with caution.

Further improvements over the {RK} algorithm have recently been obtained using
\emph{adaptive sampling}, leading to the so-called ABRA~\cite{riondato2018abra}
(by Riondato and Upfal) and KADABRA~\cite{DBLP:conf/esa/BorassiN16} (by Borassi and Natale)
algorithms. Borassi and Natale show in their paper that KADABRA dominates ABRA 
(and thus other BC approximation algorithms)
in terms of running time and approximation quality. Hence, KADABRA is the state of the art in terms
of (sequential) approximation (also see Matta \etal~\cite{Matta2019} for this conclusion). Since KADABRA is the
basis of our work, we explain it in some detail in Section~\ref{sub:kadabra}.

Surprisingly, only few works have considered parallelism in connection with approximation explicitly so far
(among them our own~\cite{DBLP:conf/europar/GrintenAM19} and to some extent Ref.~\cite{DBLP:conf/esa/BorassiN16}; both will be described later).
Hoang \etal~\cite{hoang2019round} provide ideas on how to use approximation in a distributed setting, 
but their focus is on exact computation. To the best of our knowledge, there are no MPI-based parallelizations
of KADABRA in the literature.

%========================================================================================

\section{Preliminaries}

Throughout this paper, all graphs that we consider are undirected and 
unweighted.\footnote{The parallelization
	techniques considered in this paper also apply to directed and/or weighted graphs
	if the required modifications to the underlying sampling algorithm are done.
	For more details, see Ref.~\cite{DBLP:conf/esa/BorassiN16}.}
We consider an execution environment consisting of $P$ processes
(distributed over multiple compute nodes)
and $T$ threads per process. In descriptions of algorithms, we denote the index of the
current process by $p \in \{0, \ldots, P - 1\}$;
the index of the current thread is denoted by $t \in \{0, \ldots, T - 1\}$.
Process zero ($p = 0$) and thread zero ($t = 0$) of each process sometimes
have special roles in our algorithms.

The current state-of-the-art algorithm for shared-memory parallel betweenness approximation
algorithm was presented by van der Grinten \etal~\cite{DBLP:conf/europar/GrintenAM19}.
This algorithm is a parallelization of the KADABRA algorithm by
Borassi and Natale~\cite{DBLP:conf/esa/BorassiN16}.
As our MPI-based betweenness approximation algorithm builds upon KADABRA
as the underlying sampling algorithm,
we revisit the basic ideas of KADABRA next.

\subsection{The KADABRA Algorithm}
\label{sub:kadabra}
An in-depth discussion of the KADABRA algorithm is beyond the scope
of this section; for details, we refer the reader to the
original paper~\cite{DBLP:conf/esa/BorassiN16}.
Specifically, below we do not discuss how certain functions and constants
are determined, as those computations are quite involved and not instructive
for parallelization purposes.
Like previous betweenness approximation algorithms
(such as the RK algorithm~\cite{riondato2016fast}), KADABRA samples
pairs $(s, t) \in V \times V$ of vertices $s \neq t$;
for each pair, it samples a shortest $s$-$t$ path.
From these data, the algorithm computes the number $\bccnt(x)$ of
sampled $s$-$t$ paths that contain a given vertex $x$,
for each $x \in V$. Let $\tau$ denote the number of samples
taken so far. After termination, $\bcapprox(x) := \bccnt(x) / \tau$
represents the (approximate) betweenness centrality of $x$.

KADABRA improves upon previous betweenness approximation algorithms
in two respects:
(i) it uses adaptive sampling instead of taking a fixed number of samples,
and (ii) it takes samples using a bidirectional BFS instead of an \enquote{ordinary}
(\ie unidirectional) BFS.
To check the stopping condition of KADABRA's adaptive sampling
procedure, one has to verify
whether two functions
$f(\bcapprox(x), \delta_L(x), \omega, \tau)$
and $g(\bcapprox(x), \delta_U(x), \omega, \tau)$
simultaneously assume values smaller than $\epsilon$, for all vertices $x \in V$ of the graph.
Here, $\omega$ is a statically computed maximal number of samples,
and $\delta_L(x)$ and $\delta_U(x)$ are per-node failure probabilities
computed such that
$\delta_L(x) + \delta_U(x) < \delta$ holds.\footnote{The exact choices
	for $\delta_L(x)$ and $\delta_U(x)$ do not affect the algorithm's correctness,
	but they do affect its running time.}
The constants $\delta_L$, $\delta_U$ and $\omega$ need to
be precomputed before adaptive sampling is done.
As a result, KADABRA consists of the following phases:
\begin{enumerate}
	\item \textbf{Diameter} computation. This is the main ingredient
		required to compute $\omega$.
	\item \textbf{Calibration} of $\delta_L$ and $\delta_U$.
		In this phase, the algorithm takes a few samples (non-adaptively)
		and optimizes $\delta_L$ and $\delta_U$ based on those
		initial samples.
	\item \textbf{Adaptive sampling}. The adaptive sampling phase
		consumes the majority of the algorithm's running time.
\end{enumerate}

\subsection{Parallelization of Adaptive Sampling Algorithms}

In this work, we focus on the parallelization of the adaptive sampling
phase of KADABRA.
As mentioned in the introduction, the main challenge of parallelizing adaptive sampling
is to reduce the communication overhead despite the fact that checking the stopping
condition requires global synchronization.
It is highly important that we overlap sampling
and the aggregation of sampling states: in our experiments,
this aggregation can incur a communication volume of up to 25 GiB,
while taking a single sample can be done in less than 10 milliseconds.
Thus, we want to let each thread take its own samples
independently of the other threads or processes.
Each thread $t$ conceptually updates its own $\bccnt_t$ vector
and its number of samples $\tau_t$ after taking a sample.
Checking the stopping condition requires the aggregation of all $\bccnt_t$
vectors to a single $\bccnt = \sum \bccnt_t$,
\ie the aggregation of $\Oh(P T)$ vectors
of size $\Oh(|V|)$.
Our algorithms will not maintain a ($\tau_t$, $\bccnt_t$)
pair explicitly. Instead, our algorithms sometimes have to manage multiple
$(\tau, \bccnt)$ pairs per thread. For this purpose, we call
such a pair $S := (\tau, \bccnt)$ a \emph{state frame} (SF).
The state frames comprise the entire sampling state of the algorithm
(aside from the constants mentioned in Section~\ref{sub:kadabra}).

Also note that the functions $f$ and $g$ involved in the stopping condition
are not monotone w.r.t. $\bccnt$ and $\tau$. In particular, it is not enough to simply
check the stopping condition while other threads concurrently modify
the same state frame; the stopping condition must be checked on
a consistent sampling state.
Furthermore, it is worth noting that \enquote{simple} parallelization techniques
-- such as taking a fixed number of samples before each check of the stopping
condition -- are not enough~\cite{DBLP:conf/europar/GrintenAM19}. Since they fail
to overlap computation and aggregation, they are known to not scale well,
even on shared-memory machines.

We remark that the challenges of parallelizing KADABRA and other adaptive sampling
algorithms are mostly identical. Hence, we expect that our parallelization techniques can
be adapted to other adaptive sampling algorithms easily as well.

%========================================================================================

\section{MPI-based Adaptive Sampling}
\label{sec:algorithms}

\begin{algorithm}[t]
\caption{MPI parallelization (no multithreading)}
\label{algo:mpionly}
\begin{algorithmic}[1]
	\State $S \gets \mathbf{0}$
		\Comment{aggregated state frame}
	\State $d \gets \mathsf{false}$
		\Comment{$d$ $\widehat{=}$ termination flag}
	\State $S_\loc \gets \mathbf{0}$
		\Comment{$S_\loc$ $\widehat{=}$ state frame $(\tau_p, \bccnt_p)$ of process $p$}
	\While{\textbf{not} $d$}
		\For{$n_0$ times}
				\label{line:n_zero}
				\Comment{$n_0$ $\widehat{=}$ appropriately chosen const.}
			\State $S_\loc \gets S_\loc + \Call{sample}{\null}$
		\EndFor
		\State $S_\loc' \gets S_\loc$
			\label{line:snapshot}
			\Comment{take snapshot before reduction}
		\State $S_\loc \gets \mathbf{0}$
		\State \texttt{//} aggregate all $S_\loc'$ into $S'$ at $p = 0$
		\While{$\Call{\texttt{ireduce}}{S', S_\loc'}$ is not done}
				\label{line:agg_overlap}
			\State $S_\loc \gets S_\loc + \Call{sample}{\null}$
		\EndWhile
		\If{$p = 0$}
				\label{line:check_stop}
				\Comment{only $p = 0$ checks stopping condition}
			\State $S \gets S + S'$
				\Comment{aggregate snapshot $S'$ into $S$}
			\State $d \gets \Call{checkForStop}{S}$
		\EndIf
		\State \texttt{//} send $d$ at $p = 0$, receive $d$ at $p \neq 0$
		\While{$\Call{\texttt{ibroadcast}}{d}$ is not done}
				\label{line:bcast_overlap}
			\State $S_\loc \gets S_\loc + \Call{sample}{\null}$
		\EndWhile
	\EndWhile
\end{algorithmic}
\end{algorithm}

As the goal of this paper is an efficient MPI-based parallelization
of the adaptive sampling phase of KADABRA, we need
an efficient strategy to perform the global aggregation of sampling
states while overlapping communication and computation.
MPI
provides tools to aggregate data from different processes (\ie \texttt{MPI\_Reduce})
out-of-the-box.
We can overlap communication and computation simply by
using the non-blocking variant of this collective function (\ie \texttt{MPI\_Ireduce}).
Indeed, our final algorithm will make use of MPI reductions
to perform aggregation across
processes on different compute nodes, but a more sophisticated strategy
will be required to also support multithreading.
If we disregard multithreading for a moment and rely purely
on MPI for communication, it is not too hard to construct
a parallel algorithm for adaptive sampling.
Algorithm~\ref{algo:mpionly} depicts such an
MPI-based parallelization of the adaptive sampling phase of KADABRA;
the same strategy can be adapted to other adaptive sampling algorithms.
The algorithm overlaps communication with computation by taking
additional samples during aggregation and broadcasting operations
(lines~\ref{line:agg_overlap} and~\ref{line:bcast_overlap}).
To avoid modifying the communication buffer during sampling operations
that overlap the MPI reduction, it has to
take a snapshot of the sampling state before initiating the reduction
(line~\ref{line:snapshot}).
Note that the stopping condition is only checked by a single process.
This approach is chosen so as to avoid any additional communication
-- and evaluating the stopping condition is indeed cheaper than
the aggregation required for the check
(this is confirmed by our experiments, see Section~\ref{sec:experiments}).
In the algorithm, the number $n_0$ of samples before each aggregation
(line~\ref{line:n_zero}) should be tuned in order to
check the stopping condition neither too rarely nor too often.
The less often the stopping condition is checked, the
larger the latency becomes between the point in time when the stopping
condition would be fulfilled and the point in time when the algorithm
terminates. Nevertheless, checking it too often incurs high
communication costs. We refer to Section~\ref{sub:epochlength}
for the selection of $n_0$.

Given a high-quality implementation of \texttt{MPI\_Ireduce},
Algorithm~\ref{algo:mpionly} can be expected to scale reasonably
well with the number of MPI processes, but it does not efficiently
utilize modern multicore CPUs.\footnote{We remark that,
	at least in our experiments, implementations of
	\texttt{MPI\_Ireduce} did
	\emph{not} deliver the desired performance
	(especially when compared to \texttt{MPI\_Reduce}), see Section~\ref{sub:impldetails}.}
While it is of course possible to start
multiple MPI processes per compute node (\eg one process per core),
it is usually more efficient to communicate directly via shared-memory
than to invoke the MPI library.\footnote{High-quality MPI libraries
	do implement local communication via shared-memory; still, the
	MPI calls involve some additional overhead.}
More critically, starting multiple processes per compute node limits the
amount of memory that is available to each process. As our assumption is
that each thread can take samples individually, each thread needs access
to the entire graph -- and the largest interesting graphs often fill
the majority of the total memory available on our compute nodes.
Because the graph is constant during the algorithm's execution,
sharing the graph data structure
among multiple threads on the same compute node thus allows an algorithm
to scale to much larger graphs.

In the remainder of this work, we focus on combining the basic MPI-based
Algorithm~\ref{algo:mpionly} with an efficient method to aggregate
samples from multiple threads running inside the same MPI process.
We remark that \texttt{MPI\_Ireduce} cannot be used for this purpose;
MPI only allows communication between different processes
and not between threads of the same process.
Existing fork-join based multithreading frameworks like OpenMP
do provide tools to aggregate data from different threads
(\ie OpenMP \texttt{\#pragma omp parallel for reduction});
however, they do not allow overlapped aggregation and computation.
Instead, we will use an efficient concurrent data structure
that we developed in Ref.~\cite{DBLP:conf/europar/GrintenAM19}
for the shared-memory parallelization of adaptive sampling algorithms.

\subsection{The Epoch-based Framework}

In Ref.~\cite{DBLP:conf/europar/GrintenAM19}, we presented our epoch-based framework,
an efficient strategy
to perform the aggregation of sampling states
with overlapping computation on shared-memory architectures.
In this section, we summarize the main results of that paper.
Subsequently, in Section~\ref{sub:epochbarrier} we give a functional description
of the mechanism behind the epoch-based framework without diving into the
implementation details of Ref.~\cite{DBLP:conf/europar/GrintenAM19}.
In Section~\ref{sub:epochmpi}, we construct a new MPI-based parallelization
based on the epoch-based framework.

The progress of a sampling algorithm derived from the epoch-based
framework is divided into discrete \emph{epochs}. 
The epochs are \emph{not}
synchronized among threads,
\ie while thread $t$ is in epoch $e$, another thread $t' \neq t$
can be in epoch $e' \neq e$.
Each thread $t$ allocates a new
state frame $S_t^e$ whenever it transitions to a new
epoch $e$.
During the epoch,
thread $t$ only writes to SF $S_t^e$.
Thread $t = 0$ has a special
role: in addition to taking samples, it is also responsible for
checking the stopping condition. The stopping condition is
checked once per epoch, taking into account the SFs
$S_t^e$ generated during
that epoch. To initiate a check of the stopping condition for
epoch $e$, thread zero has the ability to command all threads to advance to
an epoch $e' > e$. Before performing the check, thread zero waits for
all threads to complete the transition.
As the algorithm guarantees
that those SFs will never be written to,
this check of the stopping condition yields a sound result.

The key feature of the epoch-based framework is that
it can be implemented without introducing synchronization barriers into the
sampling threads, \ie it is wait-free for the sampling threads.
It can be implemented without the use of heavyweight synchronization
instructions like compare-and-swap (in favor of lightweight memory fences).
Furthermore, even for thread zero (that also has to check the stopping condition),
the entire synchronization mechanism can be fully overlapped
with computation.
For further details on the implementation of the epoch mechanism
(\eg details on the memory fences required for its correctness),
we refer to Ref.~\cite{DBLP:conf/europar/GrintenAM19}.

\begin{algorithm}[t]
\caption{Epoch-based MPI parallelization}
\label{algo:epochmpi}
\begin{algorithmic}[1]
	\State $S \gets \mathbf{0}$
		\Comment{aggregated state frame}
	\State $d \gets \mathsf{false}$
			\Comment{atomic termination flag}
	\State $S_t^e \gets \mathbf{0}$ for all $t$ and $e$
			\Comment{state frames}
	\State $e \gets 0$ \Comment{thread-local epoch variable}
	\If{$t \neq 0$}
		\While{\textbf{not} $d.\Call{\textnormal{\textbf{atomic\_load}}}{\null}$}
				\label{line:t_nonzero_termination}
			\State $S_t^e \gets S_t^e + \Call{sample}{\null}$
			\If{$\Call{checkTransition}{e}$}
					\label{line:epoch_check}
				\State $e \gets e + 1$
			\EndIf
		\EndWhile
	\Else\ \texttt{//} $t = 0$
		\Loop
			\For{$n_0$ times}
				\State $S_0^e \gets S_0^e + \Call{sample}{\null}$
			\EndFor
			\While{$\Call{forceTransition}{e}$ is not done}
				\label{line:epoch_force}
				\State $S_0^{e+1} \gets S_0^{e+1} + \Call{sample}{\null}$
					\label{line:nextepoch_transition}
			\EndWhile
			\State $S_\loc^e \gets \mathbf{0}$
				\Comment{aggregate epoch snapshot $S^e$ from $S_t^e$}
			\For{$i \in \{1, \ldots, T\}$}
					\label{line:epoch_agg}
				\State $S_\loc^e \gets S_\loc^e + S_i^e$
			\EndFor
			\State \texttt{//} aggregate all $S_\loc^e$ into $S^e$ at $p = 0$
			\While{$\Call{\texttt{ireduce}}{S^e, S_\loc^e}$ is not done}
				\State $S_0^{e+1} \gets S_0^{e+1} + \Call{sample}{\null}$
					\label{line:nextepoch_reduce}
			\EndWhile
			\If{$p = 0$}
				\State $S \gets S + S^e$
				\State $d' \gets \Call{checkForStop}{S}$
			\EndIf
			\State \texttt{//} send $d'$ at $p = 0$, receive $d'$ at $p \neq 0$
			\While{$\Call{\texttt{ibroadcast}}{d'}$ is not done}
				\State $S_0^{e+1} \gets S_0^{e+1} + \Call{sample}{\null}$
					\label{line:nextepoch_bcast}
			\EndWhile
			\If{$d'$} \Comment{stop threads $t \neq 0$}
				\State $d.\Call{\textnormal{\textbf{atomic\_store}}}{\mathsf{true}}$
				\State \textbf{break}
			\EndIf
			\State $e \gets e + 1$
		\EndLoop
	\EndIf
\end{algorithmic}
\end{algorithm}

\subsection{The Epoch Mechanism as a Barrier}
\label{sub:epochbarrier}

\begin{figure}[t]
\centering
\begin{tikzpicture}
	\footnotesize
	\node[minimum height=.5cm] (t0) at (0, 0) {$t = 0$};
	\node[minimum height=.5cm,below=.1cm of t0] (t1) {$t = 1$};
	\node[minimum height=.5cm,below=.1cm of t1] (t2) {$t = 2$};

	\coordinate[above right=0mm and 2mm of t0] (lt0);
	\coordinate[below right=0mm and 2mm of t0] (lb0);
	\coordinate[right=6cm of lt0] (rt0);
	\coordinate[right=6cm of lb0] (rb0);
	%\draw (lt0) rectangle (rb0);
	\coordinate[above right=0mm and 2mm of t1] (lt1);
	\coordinate[below right=0mm and 2mm of t1] (lb1);
	\coordinate[right=6cm of lt1] (rt1);
	\coordinate[right=6cm of lb1] (rb1);
	%\draw (lt1) rectangle (rb1);
	\coordinate[above right=0mm and 2mm of t2] (lt2);
	\coordinate[below right=0mm and 2mm of t2] (lb2);
	\coordinate[right=6cm of lt2] (rt2);
	\coordinate[right=6cm of lb2] (rb2);
	%\draw (lt2) rectangle (rb2);

	\coordinate[right=.5cm of lt0] (xt1);
	\coordinate[right=.5cm of lb0] (xb1);
	\coordinate[right=.5cm of lb2] (xe1);

	\coordinate[right=2cm of lt2] (xt2);
	\coordinate[right=2cm of lb2] (xb2);

	\coordinate[right=5.5cm of lt1] (xt3);
	\coordinate[right=5.5cm of lb1] (xb3);
	\coordinate[right=5.5cm of lt0] (xe3);
	\coordinate[right=5.5cm of lb2] (xf3);

	%\draw[color=orange,pattern=north west lines,pattern color=orange] (xb1) rectangle (xe3);

	% First epoch.
	\draw[color=orange,pattern=north west lines,pattern color=orange] (lt0) rectangle (xb1);
	\draw[color=orange,pattern=north west lines,pattern color=orange] (lt1) rectangle (xb3);
	\draw[color=orange,pattern=north west lines,pattern color=orange] (lt2) rectangle (xb2);

	% Second epoch.
	\draw[color=blue,pattern=north west lines,pattern color=blue] (xt1) rectangle (rb0);
	\draw[color=blue,pattern=north west lines,pattern color=blue] (xt3) rectangle (rb1);
	\draw[color=blue,pattern=north west lines,pattern color=blue] (xt2) rectangle (rb2);

	% t = 0 barrier
	\draw[color=black,line width=2mm] (xt1) -- (xb1);
	% t = 1 barrier
	\draw[color=black,line width=2mm] (xt2) -- (xb2);
	% t = 2 barrier
	\draw[color=black,line width=2mm] (xt3) -- (xb3);

	% Global barriers
	\coordinate[above=0mm of xt1] (gt1);
	\coordinate[below=0mm of xe1] (gb1);
	\coordinate[above=0mm of xe3] (gt2);
	\coordinate[below=0mm of xf3] (gb2);
	\draw[color=black,line width=.3mm] (gt1) -- (gb1);
	\draw[color=black,line width=.3mm] (gt2) -- (gb2);

	\node[above=5mm of gt1,align=center] {$\Call{forceTransition}{e}$\\called initially};
	\node[above=5mm of gt2,align=center] {$\Call{forceTransition}{e}$\\is done};
	\node[below=2mm of xb2,align=center] {$\Call{checkTransition}{e}$\\in $t = 1$};
	\node[below=2mm of gb2,align=center] {$\Call{checkTransition}{e}$\\in $t = 2$};

	\coordinate[above=2.5mm of gt1] (al);
	\coordinate[above=2.5mm of gt2] (ar);
	\path[draw,|-|,line width=.3mm] (al) -- node [above,midway] {Transition in progress} (ar);
\end{tikzpicture}
\caption{Epoch transition with $T = 3$. Orange regions: thread samples to SF of epoch $e$.
	Blue regions: thread samples to SF of epoch $(e + 1)$.
	Thick black bars: calls to \textsf{forceTransition} ($t = 0$) or
	\textsf{checkTransition} ($t \neq 0$).
	An epoch transition is always initiated by a call to \textsf{forceTransition}
	in thread zero. It always terminates with the last call to \textsf{checkTransition}
	in any other thread. The transition period (marked interval)
	is overlapped with computation (\ie sampling)
	in all threads, particularly also in thread zero.}
\label{fig:transition}
\end{figure}
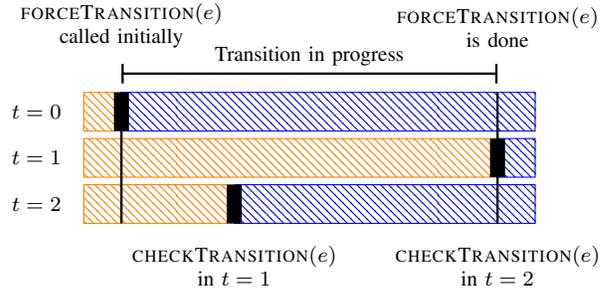

In Ref.~\cite{DBLP:conf/europar/GrintenAM19}, the epoch mechanism
was stated as a concurrent algorithm in the language
of atomic operations and memory fences.
Here, we reformulate it in a functional way;
this allows us to employ the mechanism in our MPI algorithm
without dealing with the low-level implementation details.

In our functional description, the epoch-based framework can be
seen as a specialized non-blocking barrier.
With each thread $t$, we implicitly
associate a \emph{current epoch},
identified by an integer.
We define two functions:
\begin{itemize}
\item $\Call{forceTransition}{e}$. Must only be called
	by $t = 0$ in epoch $e$.
	Initiates an \emph{epoch transition}
	and immediately causes thread zero to advance to epoch $(e + 1)$.
	This function is non-blocking, \ie
	thread zero can monitor whether
	the transition is already completed or not (in pseudocodes,
	we can treat it similarly to $\textsc{\texttt{ireduce}}$
	or $\textsc{\texttt{ibroadcast}}$).
	The initial call completes in $\Oh(1)$ steps.
	Monitoring the transition incurs $\Oh(T)$ operations per call.
\item $\Call{checkTransition}{e}$. Must only be called
	by $t \neq 0$ in epoch $e$.
	If $\Call{forceTransition}{e}$ has already been called
	by thread zero,
	\textsc{checkTransition} causes thread $t$ to participate in
	the current epoch transition.
	In this case, thread $t$ advances to epoch $(e + 1)$
	and the function returns $\mathsf{true}$.
	Otherwise, this function does nothing and returns $\mathsf{false}$.
	Completes in $\Oh(1)$ operations.
\end{itemize}

Once an epoch transition is initiated by a call to
\textsc{forceTransition} in thread zero, it remains in progress until
\emph{all} other threads perform a \textsc{checkTransition} call.
This interaction between the two functions is depicted
in Figure~\ref{fig:transition}.
Note that even during the transition (marked interval in Figure~\ref{fig:transition}),
all threads (including thread zero) can perform overlapping computation.
We remark that the epoch mechanism cannot easily be simulated
by a single generic (blocking or non-blocking) barrier since it is asymmetric:
calls to $\Call{checkTransition}{e}$ before the corresponding
$\Call{forceTransition}{e}$ have no effect
-- the calling threads do not enter an epoch transition.

\subsection{Epoch-based MPI Parallelization}
\label{sub:epochmpi}

We now describe how we combine the MPI-based approach
of Algorithm~\ref{algo:mpionly} and our epoch-based framework.
The main idea is that we use the epoch-based framework
to aggregate state frames from different threads inside the same
process, while we use the MPI-based approach of Algorithm~\ref{algo:mpionly}
to aggregate state frames among different processes.
This allows us to overlap sampling and communication both
for in-process and for inter-process communication.
Algorithm~\ref{algo:epochmpi} shows the pseudocode of the combined
algorithm.
The main difference compared to Algorithm~\ref{algo:mpionly}
is that we now have to consider multiple threads. In each process,
all threads except for
thread zero iteratively compute samples. They need to check
for epoch transitions and termination (lines~\ref{line:epoch_check}
and~\ref{line:t_nonzero_termination}) but they are not involved
in any communication or aggregation.
Thread $t = 0$ of each process proceeds in a way similar to Algorithm~\ref{algo:mpionly}; however,
it also has to command all other threads to perform an epoch transition
(line~\ref{line:epoch_force}).
After the transition is done, the state frames for
the completed epoch $e$ are aggregated locally. The result is then aggregated using
MPI. Note that after the epoch transition is initiated, thread $t = 0$
stores additional samples to the state frame for the next epoch $(e + 1)$
(lines~\ref{line:nextepoch_transition}, \ref{line:nextepoch_reduce}
and~\ref{line:nextepoch_bcast}),
so that they are properly taken into account in the next communication round.

Note that in Algorithm~\ref{algo:mpionly}, we do not synchronize the
end of each epoch across processes:
thread $t = 0$ of each process decides when to end
epochs in its process independently from all other processes. Nevertheless,
because the MPI
reduction acts as a non-blocking barrier, the epoch numbers in different
processes cannot differ by more than one.
Due to the construction of the algorithm, it is guaranteed that no thread accesses state frames of epoch $e - 2$ anymore (\ie not even thread zero).
Hence, those state frames can be reused and the algorithm only allocates two state frames
per thread.

\subsection{Length of Epochs}
\label{sub:epochlength}

The parameter $n_0$ in Algorithm~\ref{algo:epochmpi} can be tuned to
manipulate the length of an epoch. This effectively also determines
how often the stopping condition is checked.
As mentioned in the beginning of Section~\ref{sec:algorithms},
care must be taken to check the stopping condition neither too
rarely (to avoid a high latency until the algorithm terminates)
nor too often (to avoid unnecessary computation).
As adding more processes to our algorithm increases the number of
samples per epoch, we want to decrease the length of an epoch
with the number of processes. This was already observed
in Ref.~\cite{DBLP:conf/europar/GrintenAM19} -- in that paper,
we suggest to pick $n_0 = \frac{1000}{T^{1.33}}$
for a shared-memory algorithm with $T$ threads.
Both the base constant of 1000 samples per epoch and the exponent
were determined by parameter tuning.
As our MPI parallelization runs on $(P T)$ threads in total,
we adapt this number to $n_0 = \frac{1000}{(PT)^{1.33}}$.

\subsection{Accelerating Sampling on NUMA Architectures}

In preliminary experiments, we discovered that if the compute nodes'
architecture exhibits non-uniform memory access (NUMA),
it is considerably better to launch one MPI process
per socket (\ie NUMA node) instead of launching one process per compute node.
Depending on the input graph, this strategy gave a speedup
of 20-30\% on a single compute node.
This effect is due to the fact that during sampling, the algorithm
performs a significant number of random accesses to the
graph data structure (recall that each sample
requires a BFS through the graph). Launching one MPI process per
NUMA node forces the graph data structure to be allocated 
in memory that is close to the NUMA node; this decreases the
latency of cache misses to the graph
significantly.\footnote{An alternative that also benefits from
	better NUMA locality would involve the duplication of the graph
	data structure inside each process.
	We chose not to implement this solution due to higher implementation
	complexity; we expect that the difference in performance between the
	two alternatives is minimal.}
Note that launching more than one process per compute node obviously
reduces the memory available per process (as discussed in
the beginning of Section~\ref{sec:algorithms}).
Nevertheless, the compute nodes used in our experiments have two sockets
($\widehat{=}$ NUMA nodes)
and 96 GiB of memory per NUMA node; even if we launch two processes on our nodes,
we can fit graphs with billions of edges into their memory (including
all graphs from SNAP~\cite{DBLP:journals/tist/LeskovecS16}
and KONECT~\cite{DBLP:conf/www/Kunegis13}).

To take further advantage of this phenomenon, at each compute node,
we split the initial MPI communicator (\ie \texttt{MPI\_COMM\_WORLD})
into a \emph{local} communicator consisting of all processes on that node.
We also create a \emph{global} communicator consisting of the first process on each node.
We perform the MPI-based aggregation (\ie \texttt{MPI\_Ireduce}) only on
the global communicator. Before this aggregation, we aggregate over the
local communicator of each node. We perform the local aggregation via shared memory
using MPI's remote memory access functionality (in particular, passive target one-sided
communication).

\subsection{Implementation Details}
\label{sub:impldetails}

Our algorithm is implemented in C++. We use the graph data structure
of NetworKit~\cite{DBLP:journals/netsci/StaudtSM16},
a C++/Python framework for large-scale network analysis.\footnote{After
	this paper is published,
	we will make our code available on GitHub.}
In our experiments, NetworKit is configured to use 32-bit
vertex IDs. We remark that NetworKit stores both the graph
and its reverse/transpose to be able to efficiently compute a bidirectional
BFS.
For the epoch-based framework, we use the open-source
code available from Ref.~\cite{DBLP:conf/europar/GrintenAM19}.
Regarding MPI, since only thread $t = 0$ performs any MPI operations,
we set MPI's threading mode to \texttt{MPI\_THREAD\_FUNNELED}.

Recall that KADABRA requires the precomputation of the diameter
as well as an initial fixed number of samples to calibrate the algorithm.
We compute the diameter of the graph using a sequential
algorithm~\cite{DBLP:journals/tcs/BorassiCHKMT15}
-- especially for accuracies $\epsilon < 0.01$,
this phase of the algorithm only becomes significant
for higher numbers of compute nodes (see Section~\ref{sec:experiments}).
Parallelizing the computation of the initial fixed number of samples is
straightforward: we sample in all threads
in parallel, followed by a blocking aggregation (\ie \texttt{MPI\_Reduce}).

In preliminary experiments on the adaptive sampling phase,
we discovered that \texttt{MPI\_Ireduce}
often progresses much slowlier than \texttt{MPI\_Reduce} in common MPI implementations.
Hence, instead of using \texttt{MPI\_Ireduce}, we first perform a non-blocking barrier
(\ie \texttt{MPI\_Ibarrier}) followed by a blocking \texttt{MPI\_Reduce}.
This strategy resulted in a considerable speedup of the aggregation,
especially when the number of processes is increased. We remark that switching
to a fully blocking approach
(\ie dropping the \texttt{MPI\_Ibarrier} and performing a blocking reduction
after each epoch)
was again detrimental to performance.

%========================================================================================

\section{Experimental Evaluation}
\label{sec:experiments}
To demonstrate the effectiveness of our algorithm, we evaluate
its performance empirically on various real-world and synthetic graphs.
In all experiments, we pick $\delta = 0.1$
for the failure probability
(as in the original paper by Borassi and Natale~\cite{DBLP:conf/esa/BorassiN16}).
For the approximation error, we pick $\epsilon = 0.001$.
This setting of $\epsilon$ is an order of magnitude
more accurate than what was used in~\cite{DBLP:conf/europar/GrintenAM19}
-- as detailed in the introduction picking a small epsilon
is necessary to discover a larger fraction of the vertices with
highest betweenness centrality.
Note that a higher accurracy generally improves the parallel
scalability of the algorithm (and its shared-memory competitor) due to Amdahl's law:
the sequential parts of the algorithm (diameter computation and calibration)
are less affected by the choice of $\epsilon$ than the adaptive sampling
phase.
We run all algorithms
on a small cluster consisting of 16 compute nodes equipped with dual-socket
Intel Xeon Gold 6126 CPUs with 12 cores per socket.
We always launch our codes on all available cores per compute node
(with one application thread per core),
resulting in between 24 and 384 application threads in
each experiment.
Each compute node has 192 GiB of RAM available.
Intel OmniPath is used as interconnect. The cluster runs
CentOS 7.6; we use MPICH 3.2 as MPI library.

\subsection{Instance Selection}

As real-world instances, we select the largest
non-bipartite instances from
the popular KONECT~\cite{DBLP:conf/www/Kunegis13} repository
(which includes instances from SNAP~\cite{DBLP:journals/tist/LeskovecS16}
as well as the 9th and 10th DIMACS
Challenges~\cite{DBLP:conf/dimacs/dimacs74,DBLP:conf/dimacs/2012}).\footnote{We
	expect our algorithm to perform similarly well
	on bipartite graphs. Nevertheless, practitioners are probably
	more interested in centrality measures on graphs with identical
	semantics for all vertices.}
These graphs are all complex networks (specifically, they are
either social networks or hyperlink networks).
We also include some smaller road networks that proved to be
challenging for betweenness approximation in shared-memory~\cite{DBLP:conf/europar/GrintenAM19}
due to their high diameter (in particular, the largest of those
networks requires 14 hours of running time on a single node
at $\epsilon = 0.001$).
To simplify the comparison, all graphs were read as undirected and unweighted.
For disconnected graphs, we consider the largest connected component. The resulting
instances and their basic properties are listed in Table~\ref{tbl:instances}.

As synthetic graphs, we consider R-MAT graphs with
$(a, b, c, d)$ chosen as $(0.57, 0.19, 0.19, 0.05)$
(\ie matching the Graph500\footnote{\url{https://graph500.org/}} benchmarks)
as well as random hyperbolic graphs with
power law exponent $3$. Both models yield a power-law degree distribution.
We pick the density parameters of the models such that $|E| = 30\ |V|$,
which results in a density similar to that of our real-world
complex networks.

\begin{table}[tbh]
\caption{Real-world instances}
\label{tbl:instances}
\centering
\begin{tabular}{lrrr}
\toprule
Instance & $|V|$ & $|E|$ & Diameter \\
\midrule
roadNet-PA & \numprint{1087562} & \numprint{1541514} & \numprint{794} \\
roadNet-CA & \numprint{1957027} & \numprint{2760388} & \numprint{865} \\
dimacs9-NE & \numprint{1524453} & \numprint{3868020} & \numprint{2098} \\
orkut-links & \numprint{3072441} & \numprint{117184899} & \numprint{10} \\
dbpedia-link & \numprint{18265512} & \numprint{136535446} & \numprint{12} \\
dimacs10-uk-2002 & \numprint{18459128} & \numprint{261556721} & \numprint{45} \\
wikipedia\_link\_en & \numprint{13591759} & \numprint{437266152} & \numprint{10} \\
twitter & \numprint{41652230} & \numprint{1468365480} & \numprint{23} \\
friendster & \numprint{67492106} & \numprint{2585071391} & \numprint{38} \\
dimacs10-uk-2007-05 & \numprint{104288749} & \numprint{3293805080} & \numprint{112} \\

\bottomrule
\end{tabular}
\end{table}

\subsection{Parallel Scalability on Real-World Graphs}

\begin{figure}[tb]
\begin{subfigure}[t]{.475\columnwidth}
\includegraphics{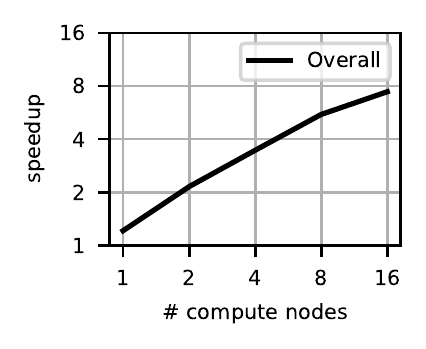}
\caption{Parallel speedup of epoch-based MPI algorithm
	over state-of-the-art shared-memory algorithm.}
\label{subfig:real-strong-su}
\end{subfigure}\hspace{0.05\columnwidth}%
\begin{subfigure}[t]{.475\columnwidth}
\includegraphics{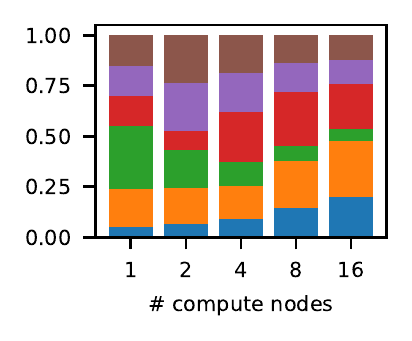}
\caption{Breakdown of running time. From bottom to top:
	diameter (blue), calibration (orange), epoch transition (green),
	non-blocking \textsc{\texttt{ibarrier}} (red),
	blocking MPI reduction (violet), check of stopping condition (brown).}
\label{subfig:real-breakdown}
\end{subfigure}
\caption{Parallel scalability on real-world graphs}
\label{fig:real-strong-scaling}
\end{figure}

\begin{figure}[tb]
\begin{subfigure}[t]{.475\columnwidth}
\includegraphics{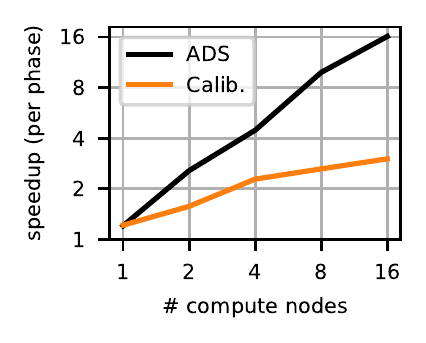}
\caption{Parallel speedup of epoch-based MPI algorithm
	over state-of-the-art shared-memory algorithm
	in adaptive sampling and calibration phases.}
\label{subfig:real-phase-su}
\end{subfigure}\hspace{0.05\columnwidth}%
\begin{subfigure}[t]{.475\columnwidth}
\includegraphics{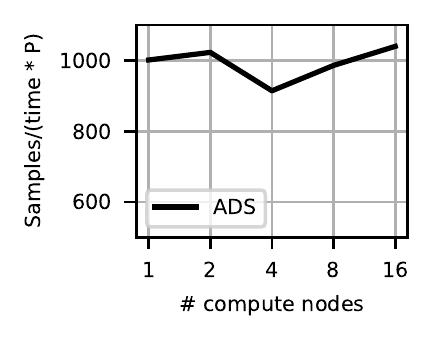}
\caption{Parallel scalability of epoch-based MPI algorithm in
	terms of samples per time and compute node.}
\label{subfig:real-tps}
\end{subfigure}
\caption{Performance characteristics on real-world-graphs}
\label{fig:real-adaptive}
\end{figure}

In a first experiment, we evaluate the performance of our
epoch-based MPI parallelization for
betweenness approximation on the real-world
instances of Table~\ref{tbl:instances}. 
In absence of prior MPI-parallel approximation algorithms
for betweenness, we compare our MPI-parallel code
with the state-of-the-art shared-memory algorithm from Ref.~\cite{DBLP:conf/europar/GrintenAM19}.
Figure~\ref{subfig:real-strong-su} depicts the speedup over this competitor.
Our MPI parallelization achives an almost linear speedup for $P \leq 8$.
For higher numbers of compute nodes, the sequential part of the computation
takes a non-negligible fraction of the running time. This can be seen from
Figure~\ref{subfig:real-breakdown}, which breaks down the fraction
of time spent in different phases of the algorithm:
indeed, the sequential diameter computation and sequential
parts of the calibration phase become more signifiant for $P \geq 8$
(blue + orange bars).
Note that the epoch transition and non-blocking barrier (green + red bars)
are overlapped communication and computation, while the aggregation (violet bar)
is the only non-overlapped communication done by the algorithm.

In Figure~\ref{subfig:real-phase-su}, we present the speedup
during the adaptive sampling and calibration phases individually.
Indeed, if we only consider the adaptive sampling phase, the algorithm
scales well to all 16 compute nodes.
While the sampling part of the calibration phase
is pleasingly parrallel, for higher numbers of compute nodes,
the sequential computations required for calibration dominate the
running time of the calibration phase.
Figure~\ref{subfig:real-tps} analyzes
the behavior during the adaptive sampling phase in more detail.
This figure demonstrates that the sampling performance scales
linearly during the adaptive sampling phase, regardless of the
number of compute nodes. This is made possible by the fact
that almost all communcation is overlapped by sampling.

\begin{table}[tb]
\caption{Per-instance statistics on 16 compute nodes,
	including number of epochs (Ep.),
	samples taken by the algorithm before termination,
	seconds spent in non-blocking \textsc{\texttt{ibarrier}} (B),
	total communication volume
	in MiB per epoch (Com.), seconds spent in adaptive sampling (Time).}
\label{tbl:n16-stats}
\centering
\begin{tabular}{lrrrrr}
\toprule
Instance & Ep. & Samples & B & Com. & Time \\
\midrule
roadNet-PA & 496 & \numprint{3943308} & \numprint{0.2} & \numprint{265.5} & \numprint{301}\\
roadNet-CA & 638 & \numprint{5269664} & \numprint{0.5} & \numprint{477.8} & \numprint{820}\\
dimacs9-NE & 79 & \numprint{669664} & \numprint{0.4} & \numprint{372.2} & \numprint{79}\\
orkut-links & 15 & \numprint{829292} & \numprint{0.2} & \numprint{750.1} & \numprint{13}\\
dbpedia-link & 11 & \numprint{1409462} & \numprint{0.3} & \numprint{4459.4} & \numprint{43}\\
dimacs10-uk-2002 & 2 & \numprint{3182023} & \numprint{8.4} & \numprint{4506.6} & \numprint{24}\\
wikipedia\_link\_en & 23 & \numprint{1129507} & \numprint{1.2} & \numprint{3318.3} & \numprint{93}\\
twitter & 26 & \numprint{1126219} & \numprint{3.3} & \numprint{10169.0} & \numprint{340}\\
friendster & 2 & \numprint{1186097} & \numprint{11.1} & \numprint{16477.6} & \numprint{50}\\
dimacs10-uk-2007-05 & 2 & \numprint{1631671} & \numprint{68.9} & \numprint{25461.1} & \numprint{184}\\

\bottomrule
\end{tabular}
\end{table}

Finally, in Table~\ref{tbl:n16-stats}, we report per-instance
statistics of the algorithm when all 16 compute nodes are used.
The road networks require the highest amount of samples but the
lowest amount communication per epoch (due to their small size).
The algorithm consequently
iterates through many epochs to solve the instance.
The largest instances (in particular \texttt{friendster} and
\texttt{dimacs10-uk-2007-05})
are solved within only two epochs since the samples
collected during the first global \textsc{\texttt{barrier}}
and aggregation are enough for KADABRA to terminate.
Again, overlapping the communication and computation
indeed allows the algorithm to be efficient
even for large communication volumes per epoch.

\subsection{Scalability \wrt Graph Size}

\begin{figure}[tb]
\begin{subfigure}[t]{.475\columnwidth}
\includegraphics{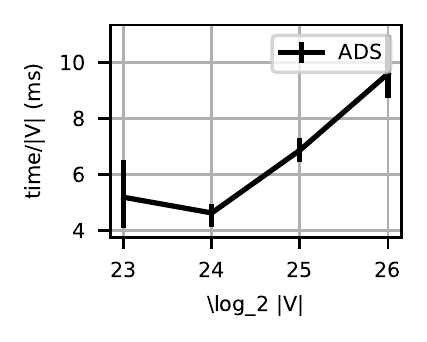}
\caption{R-MAT graphs}
\end{subfigure}\hspace{0.05\columnwidth}%
\begin{subfigure}[t]{.475\columnwidth}
\includegraphics{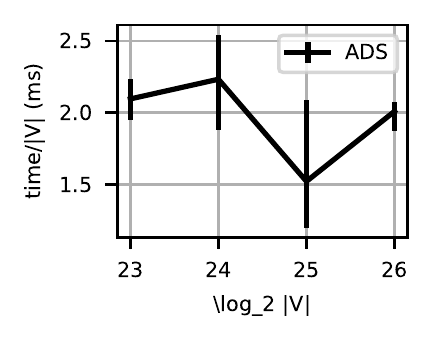}
\caption{Random hyperbolic graphs}
\end{subfigure}
\caption{Adaptive sampling time in relation to graph size on synthetic graphs}
\label{fig:synthetic}
\end{figure}

In the next experiment, we evaluate the epoch-based MPI algorithm's
ability to scale with the graph size. This experiment is performed
on synthetic R-MAT and random hyperbolic graphs. We vary the
number of vertices between $2^{23}$ and $2^{26}$
(resulting in between $250$ million and $2$ billion edges).
Figure~\ref{fig:synthetic} reports the results in terms of
time required for the adaptive sampling phase in relation to
the graph size. On R-MAT graphs, the algorithm's running time
grows slightly superlinearly: the largest graphs require $1.85 \times$
more time per vertex than the smaller graphs.
On hyperbolic graphs, the performance is mostly unaffected
by the graph size, so that we conclude that the algorithm scales 
linearly with the size of these graphs.

%========================================================================================

\section{Conclusions}
In this paper, we presented the first MPI-based parallelization of the 
state-of-the-art betweenness approximation algorithm KADABRA. 
Our parallelization is based on the epoch-based framework and non-blocking MPI collectives. 
Both techniques allow us to efficiently overlap communication and computation,
which is the key challenge for parallelizing adaptive sampling algorithms.
As a result, our algorithm is the first to allow the approximation of betweenness on
complex networks with multiple
billions of edges in less than ten minutes
at an accuracy of $\epsilon = 0.001$.

In future work, we would like to apply our method to other adaptive sampling
algorithms. As mentioned before, we expect the necessary changes to be small.

\newpage
\balance
\bibliography{dblp,biblio}

\begin{thebibliography}{10}

\bibitem{DBLP:conf/soda/AbboudGW15}
Amir Abboud, Fabrizio Grandoni, and Virginia~Vassilevska Williams.
\newblock Subcubic equivalences between graph centrality problems, {APSP} and
  diameter.
\newblock In {\em {SODA}}, pages 1681--1697. {SIAM}, 2015.

\bibitem{AlGhamdi:2017:BBC:3085504.3085510}
Ziyad AlGhamdi, Fuad Jamour, Spiros Skiadopoulos, and Panos Kalnis.
\newblock A benchmark for betweenness centrality approximation algorithms on
  large graphs.
\newblock In {\em Proceedings of the 29th International Conference on
  Scientific and Statistical Database Management}, SSDBM '17, pages 6:1--6:12,
  New York, NY, USA, 2017. ACM.

\bibitem{bader2007approximating}
David~A Bader, Shiva Kintali, Kamesh Madduri, and Milena Mihail.
\newblock Approximating betweenness centrality.
\newblock In {\em International Workshop on Algorithms and Models for the
  Web-Graph}, pages 124--137. Springer, 2007.

\bibitem{DBLP:conf/dimacs/2012}
David~A. Bader, Henning Meyerhenke, Peter Sanders, and Dorothea Wagner,
  editors.
\newblock {\em Graph Partitioning and Graph Clustering, 10th {DIMACS}
  Implementation Challenge Workshop, Georgia Institute of Technology, Atlanta,
  GA, USA, February 13-14, 2012. Proceedings}, volume 588 of {\em Contemporary
  Mathematics}. American Mathematical Society, 2013.

\bibitem{Bernaschi:2016:SBC:2903150.2903153}
Massimo Bernaschi, Giancarlo Carbone, and Flavio Vella.
\newblock Scalable betweenness centrality on multi-gpu systems.
\newblock In {\em Proceedings of the ACM International Conference on Computing
  Frontiers}, CF '16, pages 29--36, New York, NY, USA, 2016. ACM.

\bibitem{DBLP:journals/tcs/BorassiCHKMT15}
Michele Borassi, Pierluigi Crescenzi, Michel Habib, Walter~A. Kosters, Andrea
  Marino, and Frank~W. Takes.
\newblock Fast diameter and radius bfs-based computation in (weakly connected)
  real-world graphs: With an application to the six degrees of separation
  games.
\newblock {\em Theor. Comput. Sci.}, 586:59--80, 2015.

\bibitem{DBLP:conf/esa/BorassiN16}
Michele Borassi and Emanuele Natale.
\newblock {KADABRA} is an adaptive algorithm for betweenness via random
  approximation.
\newblock In {\em {ESA}}, volume~57 of {\em LIPIcs}, pages 20:1--20:18. Schloss
  Dagstuhl - Leibniz-Zentrum fuer Informatik, 2016.

\bibitem{doi:10.1080/0022250X.2001.9990249}
Ulrik Brandes.
\newblock A faster algorithm for betweenness centrality.
\newblock {\em The Journal of Mathematical Sociology}, 25(2):163--177, 2001.

\bibitem{brandes2007centrality}
Ulrik Brandes and Christian Pich.
\newblock Centrality estimation in large networks.
\newblock {\em International Journal of Bifurcation and Chaos},
  17(07):2303--2318, 2007.

\bibitem{DBLP:conf/dimacs/dimacs74}
Camil Demetrescu, Andrew~V. Goldberg, and David~S. Johnson, editors.
\newblock {\em The Shortest Path Problem, Proceedings of a {DIMACS} Workshop,
  Piscataway, New Jersey, USA, November 13-14, 2006}, volume~74 of {\em
  {DIMACS} Series in Discrete Mathematics and Theoretical Computer Science}.
  {DIMACS/AMS}, 2009.

\bibitem{geisberger2008better}
Robert Geisberger, Peter Sanders, and Dominik Schultes.
\newblock Better approximation of betweenness centrality.
\newblock In {\em Proceedings of the Meeting on Algorithm Engineering \&
  Expermiments}, pages 90--100. Society for Industrial and Applied Mathematics,
  2008.

\bibitem{hoang2019round}
Loc Hoang, Matteo Pontecorvi, Roshan Dathathri, Gurbinder Gill, Bozhi You,
  Keshav Pingali, and Vijaya Ramachandran.
\newblock A round-efficient distributed betweenness centrality algorithm.
\newblock In {\em PPoPP}, pages 272--286, 2019.

\bibitem{DBLP:conf/www/Kunegis13}
J{\'{e}}r{\^{o}}me Kunegis.
\newblock {KONECT:} the koblenz network collection.
\newblock In {\em {WWW} (Companion Volume)}, pages 1343--1350. International
  World Wide Web Conferences Steering Committee / {ACM}, 2013.

\bibitem{DBLP:journals/tist/LeskovecS16}
Jure Leskovec and Rok Sosic.
\newblock {SNAP:} {A} general-purpose network analysis and graph-mining
  library.
\newblock {\em {ACM} {TIST}}, 8(1):1:1--1:20, 2016.

\bibitem{DBLP:conf/ipps/MadduriEJBC09}
Kamesh Madduri, David Ediger, Karl Jiang, David~A. Bader, and Daniel~G.
  Chavarr{\'{\i}}a{-}Miranda.
\newblock A faster parallel algorithm and efficient multithreaded
  implementations for evaluating betweenness centrality on massive datasets.
\newblock In {\em 23rd {IEEE} International Symposium on Parallel and
  Distributed Processing, {IPDPS} 2009, Rome, Italy, May 23-29, 2009}, pages
  1--8. {IEEE}, 2009.

\bibitem{Matta2019}
John Matta, Gunes Ercal, and Koushik Sinha.
\newblock Comparing the speed and accuracy of approaches to betweenness
  centrality approximation.
\newblock {\em Computational Social Networks}, 6(1):2, Feb 2019.

\bibitem{McLaughlin:2018:AGB:3241891.3230485}
Adam McLaughlin and David~A. Bader.
\newblock Accelerating gpu betweenness centrality.
\newblock {\em Commun. ACM}, 61(8):85--92, July 2018.

\bibitem{riondato2016fast}
Matteo Riondato and Evgenios~M Kornaropoulos.
\newblock Fast approximation of betweenness centrality through sampling.
\newblock {\em Data Mining and Knowledge Discovery}, 30(2):438--475, 2016.

\bibitem{riondato2018abra}
Matteo Riondato and Eli Upfal.
\newblock Abra: Approximating betweenness centrality in static and dynamic
  graphs with rademacher averages.
\newblock {\em ACM Transactions on Knowledge Discovery from Data (TKDD)},
  12(5):61, 2018.

\bibitem{DBLP:journals/tkdd/SariyuceKSC17}
Ahmet~Erdem Sariy{\"{u}}ce, Kamer Kaya, Erik Saule, and {\"{U}}mit~V.
  {\c{C}}ataly{\"{u}}rek.
\newblock Graph manipulations for fast centrality computation.
\newblock {\em {TKDD}}, 11(3):26:1--26:25, 2017.

\bibitem{Sariyuce:2013:BCG:2458523.2458531}
Ahmet~Erdem Sariy\"{u}ce, Kamer Kaya, Erik Saule, and \"{U}mit~V.
  \c{C}ataly\"{u}rek.
\newblock Betweenness centrality on gpus and heterogeneous architectures.
\newblock In {\em Proceedings of the 6th Workshop on General Purpose Processor
  Using Graphics Processing Units}, GPGPU-6, pages 76--85, New York, NY, USA,
  2013. ACM.

\bibitem{Solomonik:2017:SBC:3126908.3126971}
Edgar Solomonik, Maciej Besta, Flavio Vella, and Torsten Hoefler.
\newblock Scaling betweenness centrality using communication-efficient sparse
  matrix multiplication.
\newblock In {\em Proceedings of the International Conference for High
  Performance Computing, Networking, Storage and Analysis}, SC '17, pages
  47:1--47:14, New York, NY, USA, 2017. ACM.

\bibitem{DBLP:journals/netsci/StaudtSM16}
Christian~L. Staudt, Aleksejs Sazonovs, and Henning Meyerhenke.
\newblock Networkit: {A} tool suite for large-scale complex network analysis.
\newblock {\em Network Science}, 4(4):508--530, 2016.

\bibitem{DBLP:conf/europar/GrintenAM19}
Alexander van~der Grinten, Eugenio Angriman, and Henning Meyerhenke.
\newblock Parallel adaptive sampling with almost no synchronization.
\newblock In {\em Euro-Par}, volume 11725 of {\em Lecture Notes in Computer
  Science}, pages 434--447. Springer, 2019.

\end{thebibliography}

\end{document}